\documentclass[prl,twocolumn,superscriptaddress,floatfix]{revtex4-1}

\usepackage{epsfig,dcolumn,amsmath,latexsym}
\usepackage{subfigure}
\usepackage[normalem]{ulem}
\usepackage{cancel}
\usepackage{multirow}
\usepackage{color}

\definecolor{red}{rgb}{1,0,0}
\definecolor{blue}{rgb}{0,0,1}
\definecolor{black}{rgb}{0,0,0}

\def\la327{La$_3$Ni$_2$O$_7$}
\def\ac327{Ac$_3$Ni$_2$O$_7$}
\def\ba21261{La$_2$BaNi$_2$O$_6$F}
\def\sr21261{La$_2$SrNi$_2$O$_6$F}

\begin{document}

\title{Ac$_3$Ni$_2$O$_7$ and La$_2Ae$Ni$_2$O$_6$F ($Ae$ = Sr, Ba): Benchmark Materials for Bilayer Nickelate Superconductivity} \preprint{1}

\author{S.-Q. Wu}
 \thanks{These authors contribute equally in this work.}
 \affiliation{School of Physics, Zhejiang University, Hangzhou 310058, China}

\author{Zihan Yang}
 \thanks{These authors contribute equally in this work.}
 \affiliation{Center of Correlated Materials, Zhejiang University, Hangzhou 310058, China}
 \affiliation{School of Physics, Zhejiang University, Hangzhou 310058, China}

\author{Xin Ma}
 \affiliation{Center of Correlated Materials, Zhejiang University, Hangzhou 310058, China}
 \affiliation{School of Physics, Zhejiang University, Hangzhou 310058, China}

\author{Jianhui Dai}
 \affiliation{Schoo of Physics, Hangzhou Normal University, Hangzhou 310036, P. R. China}

\author{Ming Shi}
 \affiliation{Center of Correlated Materials, Zhejiang University, Hangzhou 310058, China}
 \affiliation{School of Physics, Zhejiang University, Hangzhou 310058, China}

\author{H.-Q. Yuan}
 \affiliation{Center of Correlated Materials, Zhejiang University, Hangzhou 310058, China}
 \affiliation{School of Physics, Zhejiang University, Hangzhou 310058, China}

\author{Hai-Qing Lin}
 \email[E-mail address: ]{hqlin@zju.edu.cn}
 \affiliation{School of Physics, Zhejiang University, Hangzhou 310058, China}

\author{Chao Cao}
 \email[E-mail address: ]{ccao@zju.edu.cn}
 \affiliation{Center of Correlated Materials, Zhejiang University, Hangzhou 310058, China}
 \affiliation{School of Physics, Zhejiang University, Hangzhou 310058, China}
 
\date{\today}

\begin{abstract}
We theoretically propose \ac327, \ba21261, and \sr21261\ compounds to be benchmark materials for bilayer nickelate superconductivity. The stable phase of \ac327\ and \ba21261\ are found to be $I4/mmm$ without the lattice distortion caused by octahedra rotation at ambient pressure, where as the lattice distortion in \sr21261\ can be suppressed with relatively small external pressure of 4 GPa. The magnetism, electronic structure and spin susceptibilities of \ac327\ are extremely close to those of \la327\ at 30 GPa. The ground state of \ba21261\ and \sr21261\ are antiferromagnetically coupled checkerboard bilayer with sizable magnetic moment on Ni. In addition, the inter-layer coupling $J_{\perp}$ between Ni-bilayers in \ba21261\ or \sr21261\ is only $\sim$ 1/10 of that in \ac327\ or \la327\ at 30 GPa. We argue that these compounds may serve as superconducting candidates at ambient pressure and can be employed to testify theoretical proposals for bilayer nickelate superconductivity.
\end{abstract}

%\pacs{74.20.Pq, 74.70.Dd, 75.70.Tj}
\maketitle

The discovery of superconductivity near liquid-nitrogen temperature in \la327\citep{sun_signatures_2023} has drawn considerable attention from both experiment\cite{zhang_high-temperature_2023,wang_pressure-induced_2023,zhou_evidence_2023,wang_structure_2023,wang_long-range_2023,puphal_unconventional_2023,chen_polymorphism_2023} and theory\citep{luo_bilayer_2023,shen_effective_2023,liu_mathrmsifmmodepmelsetextpmfi-wave_2023,yang_possible_2023,liao_electron_2023,qin_high-t_c_2023,
luo_high-t_c_2023,wu_charge_2023,yang_minimal_2023,lu_interlayer_2023,zhang_electronic_2023,zhang_trends_2023,yang_strong_2023,
worm_spin_2023,talantsev_debye_2024,sakakibara_possible_2023,qu_roles_2023,oh_type_2023,liu_role_2023,labollita_electronic_2023,
kaneko_pair_2023,kakoi_pair_2023,jiang_high_2023,huang_impurity_2023,hou_emergence_2023,heier_competing_2023,hackner_bardasis-schrieffer-like_2023,
gu_effective_2023,geisler_structural_2023,geisler_optical_2024,christiansson_correlated_2023-1,chen_critical_2023,chen_orbital-selective_2023,cao_flat_2023}. Similar to the infinite-layer nickelate system\cite{li_superconductivity_2019}, the square Ni-O layer in \la327\ resembles the Cu-O layer in cuprates\cite{cpsr_1995_cuprate_structure}. Unlike most cuprate systems, the nickel-atoms in the two adjacent NiO layers are bonded via apical oxygen atoms, and the NiO-bilayers are separated from each other by the LaO-layer. Such bi-layer feature is found to be important in many theoretical work\cite{luo_bilayer_2023,PhysRevLett.132.036502,shen_effective_2023}. Despite of the various microscopic theoretical model study, the hallmark feature of superconductivity, zero-resistivity, appears to be sample-dependent and extremely difficult to achieve experimentally. So far, only limited groups have reported such smoking-gun evidence\cite{zhang_high-temperature_2023,wang_pressure-induced_2023}. The fragile superconductivity was attributed to the missing apical oxygen connecting Ni-atoms in adjacent bi-layers\cite{liu_mathrmsifmmodepmelsetextpmfi-wave_2023,lu_interlayer_2023,zhang_electronic_2023,labollita_electronic_2023,geisler_structural_2023,PhysRevB.109.045151,sui2023electronic,talantsev_debye_2024}.

\begin{table}
 \caption{Total energies (in meV/f.u.) of different $I4/mmm$ magnetic phases of bilayer nickelate systems. The total energies are relative to the nonmagnetic $Amam$ phase. Rows labeled with La, Ac, Ba and Sr are \la327, \ac327, \ba21261\ and \sr21261\ at ambient pressure, respectively; La-30 and Sr-4 are \la327\ at 30 GPa and \sr21261\ at 4 GPa, respectively. For \la327\ at 30 GPa, \ac327, \ba21261\ and \sr21261\ at 4 GPa, the relaxed $Amam$ structure becomes indistinguishable from the $I4/mmm$ structure. The numbers inside the brackets are magnetic moments per Ni-atom (in $\mu_B$). \label{tab:energetics}}
 \begin{tabular}{c|c|c|c|c|c}
        & NM & FM & A-AF & G-AF & C-AF \\
   \hline
 La & 55.8 &  - &  - &  - &  -  \\
 La-30 & 0.0 & -2.2 (0.1) & 0.0 (0.0) & 0.0 (0.0) & 0.0 (0.0) \\
 Ac & 0.0 & -3.3 (0.1) & 2.6 (0.3) & -0.3 (0.1) & -6.0 (0.5) \\
 Ba & 0.0 & -66.0 (0.7) & -88.5 (0.7) & -114.7 (0.7) & -132.3 (0.7) \\
 Sr & 0.4 & - &  - &  - &  -  \\
 Sr-4 & 0.0 & -34.4 (0.5) & -62.2 (0.6) & -65.9 (0.7) & -104.4 (0.7) \\
 \end{tabular}
% \begin{tabular}{c|c|c|c|c|c|c}
%        & La & La-30 & Ac & Ba & Sr & Sr-4 \\
%   \hline
% NM & 55.8 &  0.0 &  0.0 &  0.0 &  0.4 &  0.0  \\
% FM &    - & -2.2 & -3.3 &-66.0 &    - &-34.4  \\
%    &      & (0.1)& (0.1)& (0.7)&      & (0.5) \\
% A-AF &  - &  0.0 &  2.6 &-88.5 &    - &-62.2  \\
%      &    & (0.0)& (0.3)& (0.7)&      & (0.6) \\
% G-AF &  - &  0.0 & -0.3 &-114.7&    - &-65.9  \\
%      &    & (0.0)& (0.1)& (0.7)&      & (0.7) \\
% C-AF &  - &  0.0 & -6.0 &-132.3&    - &-104.4 \\
%      &    & (0.0)& (0.5)& (0.7)&    - & (0.7) \\
% $J_{\perp}$ & - & -34.4 (0.5) & -62.2 (0.6) & -65.9 (0.7) & -104.4 (0.7) \\
% \end{tabular}
\end{table}

\begin{figure*}
  \includegraphics[width=15cm]{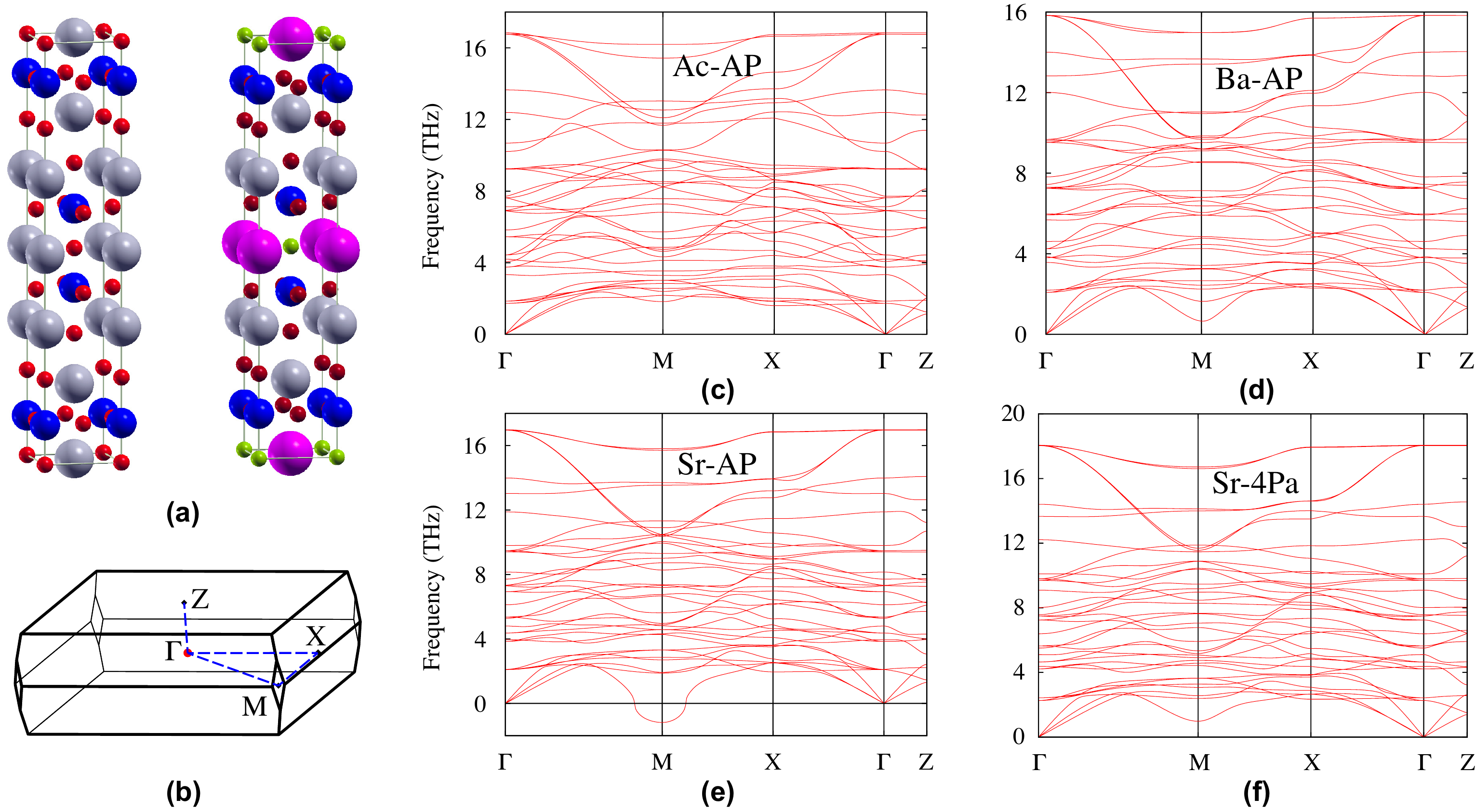}
  \caption{(a) Crystal structure of \ac327\ (left panel) and La$_2Ae$Ni$_2$O$_6$F (right panel). Blue/red spheres are Ni/O, grey spheres La, pink spheres $Ae$ (Sr or Ba), yellow spheres fluorine, respectively. (b) First Brillouine zone of $I4/mmm$ phase as well as its high symmetry path and points. (c-f) Phonon spectrum of (c) \ac327\ at ambient pressure, (d) \ba21261\ at ambient pressure, (e) \sr21261\ at ambient pressure, and (f) \sr21261\ at 4 GPa. \label{fig:geo_ph} }
\end{figure*}

One of the obstacles preventing further studies or possible applications of the bulk bilayer nickelate system is the high pressure. In \la327, the O atoms form corner-shared octahedrons with Ni atom at their centers. At the ambient pressure, the octahedrons slightly rotates, leading to an $Amam$ structure where the Ni-O$^a$-Ni angle formed by the apical oxygen atom and its nearest neighboring Ni atoms is not 180$^{\circ}$. At high pressure, such octahedral rotation is suppressed, leading to $Fmmm$ structure around 15 GPa, and the superconductivity emerges. It is widely believed that the suppression of structural distortion is crucial to the emergence of superconductivity, thus a stable $Fmmm$ phase bilayer nickelate system at lower pressure or even ambient pressure is highly desirable. Previously, there were efforts to introduce chemical pressure by replacing La by other Lanthanum elements\cite{zhang_trends_2023,sui2023electronic,geisler_structural_2023,rhodes_structural_2023}. It was found that replacing La with smaller elements cannot stabilize $Fmmm$ structure at the ambient pressure. It is worth noting that the orthohombic $Fmmm$ phase is very close to the tetragonal $I4/mmm$ phase, since its in-plane lattice constants are nearly identical ($b/a\approx 1.0136$)\cite{sun_signatures_2023}. In DFT calculations, this in-plane anisotropy is even smaller and disappears at 20 GPa\cite{geisler_structural_2023}. 

In this paper, we theoretically propose 2 bilayer nickelate compounds whose stable structures are found to be $I4/mmm$ at the ambient pressure: \ac327\ and \ba21261. In addition, we also propose \sr21261\ whose structural distortion can be suppressed at relative low pressure. Therefore, these compounds may exhibit superconductivity at ambient pressure and could serve as benchmark systems for theoretical models. In particular, \ac327\ is extremely similar to \la327\ in terms of the electronic structure, magnetism and spin fluctuations; whereas the La$_2Ae$Ni$_2$O$_6$F exhibit antiferromagnetic (AFM) ground states and reduced inter-layer coupling compared to \ac327\ or \la327\ at 30 GPa due to weaker bonding between the apical F-atoms and Ni-atoms.

There are several principles we try to follow when searching for the possible benchmark compounds. Firstly, the $I4/mmm$ phase of the resulting compounds should be energetically and dynamically stable at ambient pressure. Previous efforts have found that the chemical pressure effect by substituting La with other smaller elements indeed promotes the octahedral rotation instead of suppressing it. Therefore we try to look for elements with larger ionic radius. Secondly, the nominal valence of Ni in the resulting compounds should be the same as \la327, i.e. 2.5+. Finally, the electronic structure of the resulting compounds should be close to \la327. The second and third principle are requirements for the resulting compound to serve as benchmark system. Following these principles, we eventually came up with 3 compounds, \ac327, \ba21261, and \sr21261. \ac327\ is resultant of element substitution of La by Ac, whereas \ba21261\ and \sr21261\ can be viewed as replacing the La-O plane between the Ni-O bilayer with Ba-F and Sr-F plane, respectively (FIG. \ref{fig:geo_ph}a).

We start off by examining the energetics of these compounds (TAB. \ref{tab:energetics}). For \la327, the nonmagnetic (NM) $I4/mmm$ phase is 55.8 meV/f.u. higher than the $Amam$ phase at ambient pressure (AP-\la327); whereas the $Amam$ phase automatically relax to $I4/mmm$ phase at 30 GPa (HP-\la327). This is consistent with experimental observation\cite{sun_signatures_2023,wang_structure_2023}, as well as previous calculations\cite{geisler_structural_2023}. Similarly, for \sr21261, the $I4/mmm$ phase is only $\sim$ 0.4 eV/f.u. higher than the $Amam$ phase (AP-\sr21261), and the $Amam$ phase automatically relax to $I4/mmm$ phase at 4 GPa (LP-\sr21261). In contrast, the $I4/mmm$ phases of \ac327\ and \ba21261\ are already favored at ambient pressure. In addition, the phonon dispersion of $I4/mmm$ phase \ac327\ and \ba21261\ are stable without any imaginary phonon modes at ambient pressure [FIG. \ref{fig:geo_ph}(c-d)]. The phonon dispersion of $I4/mmm$ phase \sr21261\ exhibits small imaginary phonon modes around $M$ at ambient pressure, which is suppressed for LP-\sr21261\ [FIG. \ref{fig:geo_ph}(e-f)]. Therefore, the stable phases of \ac327\ and \ba21261\ at ambient pressure are $I4/mmm$, where as the lattice distortion of \sr21261\ can be suppressed with small external pressure. The formation energy of \sr21261\ (\ba21261) at ambient pressure is calculated to be $E_f = -$775 ($-$8.7) meV/f.u. by assuming reaction compounds to be $Ae$O, $Ae$F$_2$, LaNiO$_2$ and LaNiO$_3$ (refer to SI). The $E_f$ of \sr21261\ is almost twice the number of \la327, suggesting it is energetically very stable. Furthermore, the dissociation energy of the apical F atom in \sr21261\ (\ba21261) is also $\sim$87 kJ/mol higher than that of apical O atom in \la327\ or \ac327, thus the \sr21261\ or \ba21261\ structure is less susceptible to the loss of apical atoms (refer to SI).

\begin{figure*}
  \includegraphics[width=15cm]{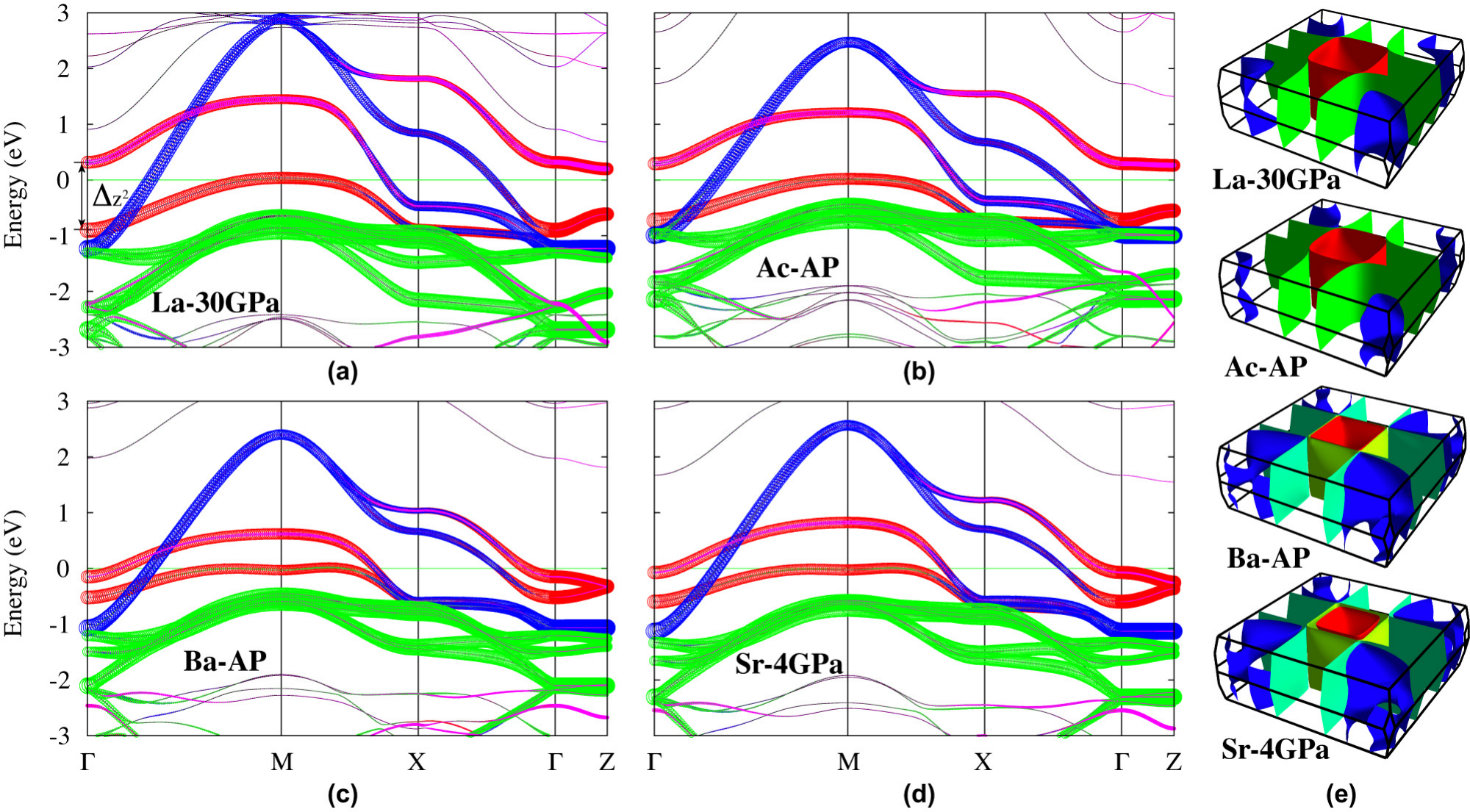}
  \caption{(a-d) Band structure and (e) fermi surfaces. La-30 GPa is \la327\ at 30 GPa, Ac-AP is \ac327\ at ambient pressure, Ba-AP is \ba21261\ at ambient pressure, and Sr-4 GPa is \sr21261\ at 4 GPa. The sizes of the red, blue, green and purple circles are proportional to the weight of Ni-3d$_{z^2}$, Ni-3d$_{x^2-y^2}$, Ni-3d$_{t2g}$, and apical O/F-2p$_z$ orbitals.\label{fig:bs} The red, green, and blue Fermi surface sheets corresponds to the $\alpha$, $\beta$, and $\gamma$ bands, respectively.}
\end{figure*}

In TAB \ref{tab:energetics}, the relative energies and magnetic moments of different magnetic configurations are also listed. We consider 4 different possible magnetic states, including ferromagnetic (FM), A-type inter-layer antiferromagnetic (A-AF), inter-layer FM with in-plane checkerboard (C-AF), and inter-layer antiferromagnetic with in-plane checkerboard (G-AF). For HP-\la327 and \ac327\ at ambient pressure, all magnetic states are degenerate with NM state within the DFT errorbar. Such behavior suggests the presence of large spin fluctuation in both cases. For \ba21261\ at ambient pressure, however, magnetic long-range order prevails with significantly lower total energy and larger magnetic moment per Ni atom. Thus, the ground state of \ba21261\ is C-AF. Such AF pattern is close to the cuprate systems\cite{RevModPhys.70.897}. Employing a Heisenberg-like model including only the nearest neighboring coupling $H^{\mathrm{mag}}=J_{\parallel}\sum_{<i,j>_{\parallel}}\mathbf{S}_i\cdot\mathbf{S}_j+J_{\perp}\sum_{<i,j>_{\perp}}\mathbf{S}_i\cdot\mathbf{S}_j$, one can estimate $J_{\parallel}S^2\approx$5.8 meV and $J_{\perp}S^2\approx$10.0 meV, respectively. Here, $<i,j>_{\parallel}$/$<i,j>_{\perp}$ indicate the in-plane/inter-layer nearest neighboring Ni-sites, and $J_{\parallel}$/$J_{\perp}$ are the respective couplings. For LP-\sr21261, its ground state is also C-AF. Compared to \ba21261, its in-plane coupling $J_{\parallel}S^2$ is reduced to 4.9 meV, whereas the inter-plane coupling $J_{\perp}S^2$ is significantly increased to 14.4 meV. The $J_{\perp}S^2$ obtained in \ba21261\ and LP-\sr21261\ are one order of magnitude smaller than the value estimated in HP-\la327\ or \ac327, reflecting reduced coupling between bi-layer systems in $Ae$-F substituted systems. 

\begin{table}
 \caption{Leading hopping parameters of 4-orbital tight-binding models (in eV). The first row is the original minimum model parameters obtained by Luo {\it et al.} in Ref. \citep{luo_bilayer_2023}. La-30 is \la327\ at 30 GPa, Ac is \ac327\ at ambient pressure, Ba is \ba21261\ at ambient pressure, Sr-4 is \sr21261\ at 4 GPa. \label{tab:hoppings}}
 \begin{tabular}{c|c|c|c|c|c|c|c|c}
       &  $t^x_1$  & $t^z_1$ & $t^x_2$ & $t^z_2$ & $t^{xz}_3$ & $t^x_{\perp}$ & $t^z_{\perp}$ & $t^{xz}_4$ \\ %& $\epsilon^x$ & $\epsilon^z$ 
   \hline\hline
 Luo   & -0.483 & -0.110 & 0.069 & -0.017 & 0.239 &  0.005 & -0.635 & -0.034 \\ %& 0.776 & 0.409  
 La-30 & -0.482 & -0.117 & 0.064 & -0.012 & 0.237 &  0.002 & -0.628 & -0.039 \\ %& 0.772 & 0.378  
 Ac    & -0.412 & -0.092 & 0.066 & -0.012 & 0.195 &  0.006 & -0.554 & -0.033 \\ %& 0.679 & 0.339  
 Ba    & -0.408 & -0.073 & 0.070 & -0.012 & 0.181 & -0.007 & -0.182 & -0.021 \\ %& 0.604 & 0.117  
 Sr-4  & -0.435 & -0.084 & 0.068 & -0.012 & 0.196 & -0.009 & -0.267 & -0.039 \\ %& 0.604 & 0.117  
 \end{tabular}
\end{table}

\begin{figure*}[htp]
  \includegraphics[width=15cm]{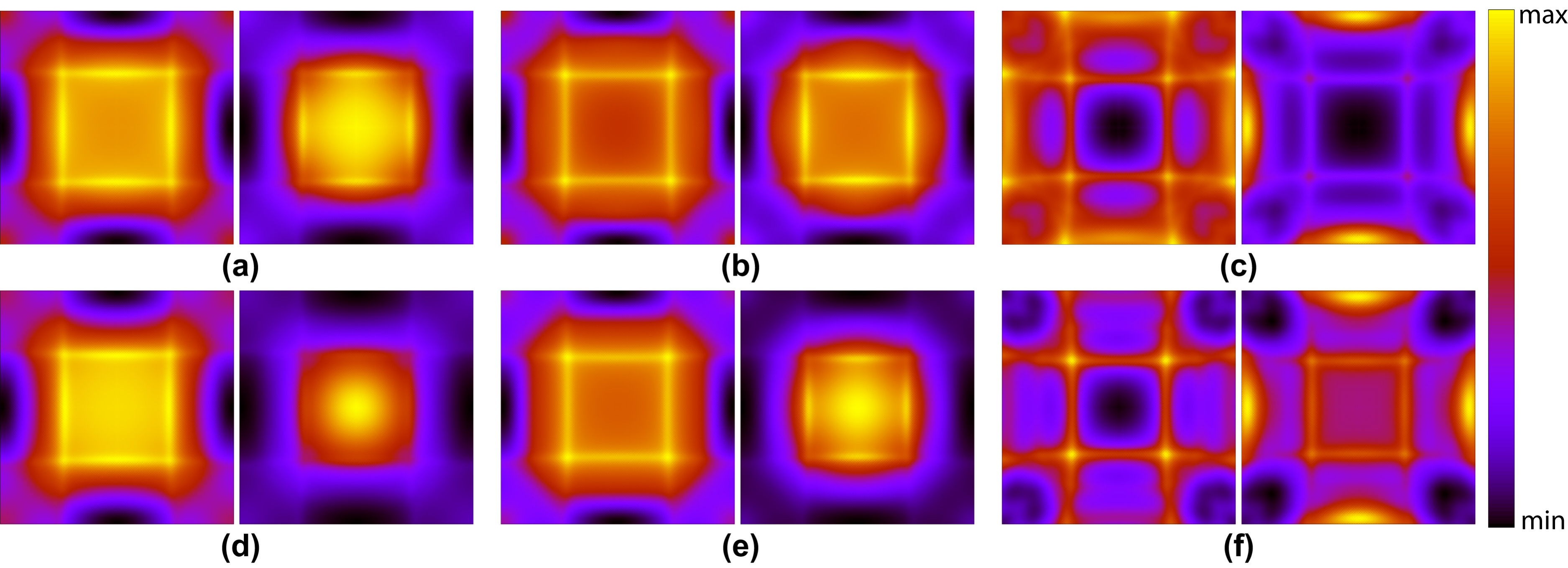}
  \caption{Bare electron susceptibility $\mathrm{Tr}[\chi^0(\mathbf{q})]$ (left panels) and spin susceptibility $\mathrm{Tr}[\chi^S(\mathbf{q})]$ (right panels) of (a-c) \la327\ at 30 GPa and (d-f) \ac327\ at ambient pressure. (a) and (d) are pristine compound without doping; (b) and (e) 0.1 hole-doped; (c) and (f) 0.2 hole-doped. \label{fig:chi} }
\end{figure*}

The band structure of these compounds in NM $I4/mmm$ phase are compared in FIG. \ref{fig:bs}. In general, the band structure of these compounds are extremely similar to one another. Compared to HP-\la327, the Ni-3d$_{x^2-y^2}$ bands of \ac327\ are less dispersive due to the increased lattice constants (refer to SI), so the band width of Ni-3d$_{x^2-y^2}$ is reduced from $\sim$ 4 eV in \la327\ to $\sim$ 3.5 eV in \ac327. However, the orbital characters of \ac327\ band structure near the Fermi level are nearly identical to those of \la327. In particular, the apical oxygen O$^a$-2p$_z$ orbitals significantly hybridizes with the anti-bonding state of Ni-3d$_{z^2}$ [FIG. \ref{fig:bs} (a,b)]. Therefore, the splitting at $\Gamma$ between bonding-antibonding state of Ni-3d$_{z^2}$ $\Delta_{z^2}\approx1.1$ eV is still large, and the antibonding state of Ni-3d$_{z^2}$ is unoccupied at $\Gamma$. Thus, the Fermi surface topology remains the same between \la327\ and \ac327. On the contrary, in both \ba21261\ and LP-\sr21261\ [FIG. \ref{fig:bs} (c-d)], the weight of apical fluorine F-2p$_z$ is nearly absent in the anti-bonding state of Ni-3d$_{z^2}$, suggesting negligible hybridization between these states. As such, $\Delta_{z^2}$ is substantially reduced to $\sim 0.37$ eV in \ba21261, and the antibonding state of Ni-3d$_{z^2}$ becomes occupied at $\Gamma$. We note that external pressure will enhance the hybridization, thus $\Delta_{z^2}$ of \sr21261\ at 4 GPa is enhanced to $\sim0.5$ eV. However, the antibonding state at $\Gamma$ remains occupied up to 8 GPa in \sr21261. 

The change of band structure is also reflected in the tight-binding model fitted with Ni-3d$_{eg}$ orbitals using maximally projected Wannier function method\cite{method:mpwf} (TAB. \ref{tab:hoppings}). Compared to the hopping parameters of HP-\la327, the in-plane nearest neighboring hoppings $t_1^x$ and $t_1^z$ are smaller in these compounds, due to expanded Ni-Ni distances. The leading hybridization between Ni-3d$_{x^2-y^2}$ and Ni-3d$_{z^2}$ $t_3^{xz}$ is also slightly reduced, but is of similar magnitude. The Ni-3d$_{z^2}$ inter-layer coupling $t^z_{\perp}$ in \ba21261\ or LP-\sr21261\ is only $\sim$ 1/3 of that in HP-\la327. As a comparison, $t^z_{\perp}$ in \ac327\ is of similar magnitude as that in HP-\la327. Assuming super-exchange mechanism via apical oxygen or fluorine where $J_{\perp}\propto t^2/U$, this implies that $J_{\perp}$ in \ac327\ or HP-\la327\ is approximately 10 times larger than those in the fluorine compounds. Such estimation suggests $J_{\perp}S^2$ is  order of $\sim$ 100 meV in \la327\ or \ac327. This is of similar magnitude as the one obtained in previous studies\cite{shen_effective_2023}.

We have also calculated the Coulomb interaction parameters of these compounds using constrained random phase approximation (cRPA) method\cite{PhysRevB.83.121101}. In such calculations, we employ Ni-3d$_{eg}$ orbitals from both Ni-sites in the primitive cell, so that the inter-site interactions $V$ between nearest neighboring Ni-atoms across the bi-layer, in addition to the on-site interactions $U$, can be obtained. In both \ac327\ and HP-\la327, the on-site $U=$2.2 eV and $J=$0.8 eV; whereas in \ba21261\ and LP-\sr21261, $U=1.6$ eV and $J=0.7$ eV. Interestingly, the inter-site $V$ is large in both \ac327\ ($\sim$0.9 eV between d$_{z^2}$) and HP-\la327 ($\sim$0.6 eV between d$_{z^2}$), but is small in \ba21261\ ($\sim$ 0.2 eV between d$_{z^2}$) and \sr21261\ at 2 GPa ($\sim$ 0.3 eV between d$_{z^2}$). Since $V$ reflects the overlap between inter-layer Wannier orbitals, it is an indication of inter-layer coupling as well. Therefore, it also suggests reduced inter-layer coupling in \ba21261\ and \sr21261.

Finally, we compare the bare electron susceptibility and spin susceptibility of HP-\la327 and \ac327\ at ambient pressures (FIG. \ref{fig:chi}). Employing the complete Wannier orbital hamiltonian which perfectly reproduces first-principles results including dispersion along $c$-axis, the full bare susceptibility matrix is calculated using
% $$\chi^{0pq}_{st}(\mathbf{q}, \mathrm{i}\nu)=-\frac{1}{N_{\mathbf{k}}\beta}\sum_{n,\mathbf{k}}G_{sp}(\mathrm{i}(\omega_n+\nu), \mathbf{k+q})G_{qt}(\mathrm{i}\omega_n, \mathbf{k})$$
$${\chi^0}^{pq}_{st}(\mathbf{q}, \mathrm{i}\nu)=-\frac{1}{N_{\mathbf{k}}\beta}\sum_{n,\mathbf{k}}G_{sp}(\mathrm{i}\omega_n, \mathbf{k})G_{qt}(\mathrm{i}\omega_n+i\nu, \mathbf{k+q})$$
and the trace is defined as $\mathrm{Tr}[\chi^0(\mathbf{q})]=\sum_{p,s}{\chi^0}^{pp}_{ss}$. The spin susceptibility $\chi^S(\mathbf{q})$ is also calculated using the full bare susceptibility matrix at the RPA level assuming atomic-like on-site interaction with $U$=0.9 eV and $J/U=0.16$. The details of the formalism are explained in the Supplementary Information. We note the chosen $U$ is well below the critical $U_c=\min_{\mathbf{q}}\lbrace U |  U\chi^0(\mathbf{q})=1\rbrace$, which signals magnetic instability\cite{PhysRevB.69.104504,PhysRevLett.108.017001}. In general, the bare and spin susceptibility of \ac327\ is very similar to HP-\la327 at all the doping levels. It is therefore reasonable to speculate ambient pressure superconductivity in \ac327. In our calculations, the complete Wannier orbital hamiltonian faithfully reproduces the DFT band structure, the $\alpha$ and $\gamma$ pockets become dispersive in $k_z$-direction. This is particularly notable in the $\Gamma$-X direction for $\gamma$-pocket. As a result of the severe dispersion, the nesting between the $\alpha$ and $\gamma$ pockets is suppressed compared to the minimum model based calculations. On the contrary, the $\gamma$-pocket remains dispersionless along $k_z$ at the BZ boundary, thus the nesting between the $\beta$ and $\gamma$ pockets is preserved. Such nesting vector is along $\Gamma$-M $(\pi, 0)$, and favors $d_{xy}$ pairing. Similar results is also obtained in Ref. \cite{liu_role_2023}. For \sr21261\ at 4 GPa and \ba21261, the critical $U_c$ is only $\sim$ 0.6 eV. This is reasonable because their ground state is C-AF in DFT calculations. 

%\begin{table}
% \caption{Key aspects of the proposed compounds and \la327\ at 30 GPa. La-30 is \la327\ at 30 GPa, Ac is \ac327\ at ambient pressure, Ba is \ba21261\ at ambient pressure, Sr-4 is \sr21261\ at 4 GPa. \label{tab:key_param}}
% \begin{tabular}{c|c|c|c|c}
%            &  La-30  & Ac & Ba & Sr-4  \\ %& $\epsilon^x$ & $\epsilon^z$ 
%   \hline\hline
%  $a$ (\AA) & 
%  $c$ (\AA) &
%$E_f$ (meV) & 
% $t^z_{\perp}$ &
% $J_{\perp}$ &
%  
% \end{tabular}
%\end{table}

To summarize, we theoretically proposed possible bench mark materials for bilayer nickelate superconductors. Their $I4/mmm$ structures are energetically and dynamically stable at ambient or small external pressures. The electronic structure, magnetic properties, as well as the bare and spin susceptibility of \ac327\ are very close to HP-\la327. In contrast, the ground state of \sr21261\ and \ba21261\ are C-AF, similar to cuprate systems. The bilayer coupling is of similar strength in \ac327, but is one order of magnitude smaller in \sr21261\ and \ba21261. The striking similarities between \ac327\ and HP-\la327 suggest it could be a candidate for bilayer nickelate superconductor at ambient pressure. In addition, if superconductivity can be achieved by suppressing magnetism in \sr21261\ or \ba21261 via doping or pressure, it would be interesting to investigate the role played by interlayer coupling, which poses extra constrain for theoretical models.

%Should superconductivity be realized in \sr21261\ by hole doping, the pairing is expected to be $s_{\pm}$ whereas the pairing of \ac327\ is expected to be $d_{xy}$ due to finite $k_z$-dispersion.

\begin{acknowledgements}
The authors would like to thank Jian-Xin Zhu, Yanming Ma, Guanghan Cao, Yang Liu, Yu Song, Lin Jiao, Yongkang Xu, and Chenchao Xu for stimulating discussions. This work was support by the National Key R\&D Program of China (No. 2022YFA1402200, 2022YFA1402701), NSFC (11874137, 11974244, 12088101, 12274364), and the Key R\&D Program of Zhejiang Province (2021C01002). The calculations were performed on clusters at the High Performance Computing Cluster at Center of Correlated Matters Zhejiang University and High Performance Computing Center at Hangzhou Normal University.
\end{acknowledgements}

\bibliography{la327}

\end{document}